# Distribution of ripples in graphene membrane

Jin Yu[1,2], Mikhail I. Katsnelson[2], and Shengjun Yuan[3,2]

1 Shanghai Key Laboratory of Mechanics in Energy Engineering, Shanghai Institute of Applied Mathematics and Mechanics, School of Mechanics and Engineering Science, Shanghai University, Shanghai, 200444, China

2 Institute for Molecules and Materials, Radboud University, Heijendaalseweg 135, 6525AJ Nijmegen, The Netherlands

3 Key Laboratory of Artificial Micro- and Nano-Structures of Ministry of Education and School of Physics and Technology, Wuhan University, Wuhan 430072, China

**Abstract**

Intrinsic ripples with various configurations and sizes were reported to affect the physical and chemical properties of 2D materials. By performing molecular dynamics simulations and theoretical analysis, we use two geometric models of the ripple shape to explore numerically the distribution of ripples in graphene membrane. We focus on the ratio of ripple height to its diameter (t/$D$) which was recently shown to be the most relevant for chemical activity of graphene membranes. Our result demonstrates that the ripple density decreases as the coefficient t/$D$ increases, in a qualitative agreement with the Boltzmann distribution derived analytically from the bending energy of the membrane. Our theoretical study provides also specific quantitative information on the ripple distribution in graphene and gives new insights applicable to other 2D materials.

**TOC**

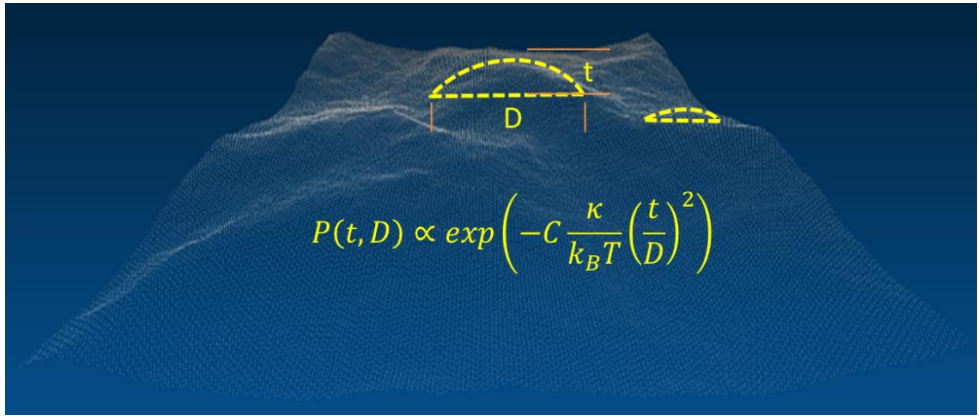

Thermally induced (Intrinsic) rippling of two-dimensional (2D) membranes is an unavoidable consequence of thermal fluctuations; the theory of structural state of 2D systems at a finite temperature is an actively developing field of statistical physics [1-9]. It was enormously stimulating by an isolation of graphene [10], experimental discovery of ripples on freely suspended graphene [11] and first atomistic simulations of ripples on graphene [12] (for the other aspects of the problem, see Refs. [13-16]). Apart from the intrinsic ripples, artificial periodic structures of ripples were introduced by utilizing the negative expansion coefficient of graphene sheet [17]. The existence of ripples affects strongly the electron transport in graphene [18], its chemical activity [19], mechanical properties [20-22] and other important characteristics. Very recently, it was demonstrated that ripples play a crucial role in hydrogen permeation through a single-layer graphene membrane [23].

The latter problem is especially interesting. According to the scenario suggested in Ref. [23], the first step in the permeation of molecular hydrogen through graphene is its dissociation of hydrogen atoms at the ripples, and the curvature of the ripples is crucially important for this process. Atomistic simulations [23] show that the enhanced chemical activity for the dissociation requires high enough curvature of the ripples, with the ratio $t/D$ larger than 0.04 (here $t$ and $D$ are the height and diameter of a ripple). The fraction of hydrogen molecular going through monolayer graphene increases when the temperature increases, and we attributed this to the increasing density of ripples with a large ratio $t/D$ [23]. Therefore, it is necessary to figure out the distribution of the ripples in their geometry at different temperatures. This is the motivation for our work where we solve this problem via molecular dynamic simulations.

Before doing so, let us make a simple estimate of the expected distribution. Let us start with the simple expression for the bending energy of the membrane [1,15,24]:

$$E = \frac{\kappa}{2} \int d^2 x (\nabla^2 h)^2 \qquad (1)$$

where $\kappa$ is the bending rigidity, $x$ is the two-dimensional vector, and $h(x)$ is the profile of the ripple. A very naive estimate of the integral in Eq. (1) gives the value of the order of $(t/D)^2$. Substituting this estimate into Boltzmann distribution $exp(-E/k_BT)$, we find a quite simple expression for the distribution function of the ripples in their geometric characteristics:

$$P(t,D) = A exp\left(-C \frac{\kappa}{k_B T} \left(\frac{t}{D}\right)^2\right) \quad (2)$$

where $C$ is an unknown numerical factor which cannot be estimated from a simple analysis. Keeping in mind that at room temperature for the case of graphene, $\frac{\kappa}{k_B T} \approx 40$, one can assume that 'chemically active' ripples should be quite well presented at room temperature, and with the temperature increase its number grows, both conclusions are in agreement with the computational data [23].

According to Eq. (2) the probability distribution of the thermal ripples depends only on the ratio $t/D$ but not on the geometric sizes ($t$ or $D$) separately. One should keep in mind however that Eq. (1) and, thus, Eq. (2) do not take into account the renormalization of bending rigidity via anharmonic coupling of in-plane and out-of-plane modes [1-9,12,15,16]. The latter takes place for the fluctuations with small enough wave vectors

$$q < q^* = \sqrt{\frac{3k_B TY}{16\pi \kappa^2}}, \quad (3)$$

where $Y$ is the two-dimensional Young modulus. For graphene at room temperature[12,15,16] $q^* \approx 2$ $(nm)^{-1}$. This means that Eq. (2) is valid only for strong enough inhomogeneities, that is, for $D < L_{th} = \frac{2\pi}{q^*}$, $L_{th} \approx 3$ $nm$ for graphene at room temperature. For larger ripples, one can expect a renormalization of the bending rigidity,

$$\kappa \to \kappa_R \approx \kappa \left(\frac{q^*}{q}\right)^\eta \approx \kappa \left(\frac{D}{L_{th}}\right)^\eta, \quad (4)$$

where $\eta \approx 0.8$ is the corresponding 'critical exponent' [1-9,12,15,16].

Thus, one can suppose that Eq. (2) is valid for small enough ripples, $D < L_{th}$, whereas for the opposite limit $D > L_{th}$, it should be replaced by the equation

$$P(t,D) = A exp\left(-C \frac{\kappa}{k_B T} \left(\frac{t}{D}\right)^2 \left(\frac{D}{L_{th}}\right)^\eta\right). \quad (5)$$

One can expect therefore that for large ripples, $D > L_{th}$, its probability decreases exponentially with the increase of the ripple size.

Now we will verify our estimation and compare these preliminary considerations with the numerical results obtained from molecular dynamics simulations.

For this and the following section, we performed a series of semiclassical molecular dynamics simulations using large-scale atomic/molecular massively parallel simulator (LAMMPS).[25] To keep it consistent with the data in Ref [23], we consider a sheet of graphene consisting of 387,200 atoms in a periodic cell of $L_x \times L_y$ = 95.4×101.1 nm$^2$ (220×440 orthorhombic unit cells) using the modified Tersoff potential. Here, the *x* and *y* directions correspond to zigzag and armchair axes, respectively. Our numerical tests confirm that the above sample size is large enough to replicate correctly the ripple-ripple correlations (see the converge tests in the Supplemental Material). The perpendicular cell size is set to be $L_z$ = 8 nm to avoid interaction between periodic images of the sheet. Meanwhile, if we choose a hexagonal super cell instead of rectangle, the distribution function becomes slightly larger, but the overall conclusion remains the same (more results with a hexagonal super cell are presented in the Supplemental Materials). Therefore, in the following, all the discussions are based on the rectangle super cell. The typical configurations at different T were obtained after thermalization with every 100,000 steps (0.000025 ps per step) and averaged over twenty of such snapshots. In Figure 1a, we show one example using a model of circle-like ripple. In our simulation, the ripples are selected by the profile of the center carbon atom which has a minimum value of 0.04 Å to its closest neighbor. By scanning each site in a snapshot obtained during the molecular dynamics simulations, we first identify all possible ripple centers as the highest carbon atoms in a circle with a radius of $R_{cut}$ = 20 Å. As for one specific ripple, the profiles could be different along different directions, thus, we introduce $\theta$ as a measure of the angle between a vector span by the radius $R$ and the armchair edge. Then one ripple is divided into 180 sectors (each in 2°) to obtain the effective radius along each direction $\theta$. In this way, all the atomic sites within a possible ripple will be taken into account. Due to the criteria of monotonicity in scanning of $R_\theta$, one ripple center will be counted only once, and will not be considered in other ripples. After collecting all the date of $\{R_\theta\}$, we can describe a ripple or characterize its size in two models: one is a circle-like model $D = 2R_{min}$ where $R_{min}=\min\{R_\theta\}$, and the other one is an arbitrary-shape model ($\{D\} = \{R_\theta + R_{\theta+180°}\}$). One should notice that, in the second model, the ripple size is not characterized by a single value but a set of $\{D\}$, which is, in principle, more accurate to describe a real profile.

In graphene, the ripple size $D$, no matter defined in the circle-like or arbitrary-shape model, is counted based on the in-plane projection between two carbon atoms and in the limit without thermal-fluctuation it should have discrete values following $D = |m\mathbf{a}+n\mathbf{b}|$ ($m$ and $n$ are integers and interatomic distances $|\mathbf{a}| = |\mathbf{b}| = 1.42$ Å). As the actual thermal-fluctuation induced displacement of a carbon atom in real space is relatively small comparing to the interatomic distances in pristine graphene, the obtained ripple sizes $D$ are distributed closely around these discrete values of $|m\mathbf{a}+n\mathbf{b}|$. Now, we make a further analysis using the circle-like model. In Figures 1b and 1c, we plot the statistical results of ripple density as a function of ripple size $D$. We see clearly that for a finite temperature, ripples with different sizes are not equally distributed in graphene membrane. The distribution of $D$ forms peaks slightly larger than those discrete values allowed in pristine graphene ($|m\mathbf{a}+n\mathbf{b}|$, indicated by the purple bar on the top). This is consistent with a common sense that thermal fluctuations induce the out-of-plane vibrations of carbon atoms, resulting in a stretching of atoms in the plane. The profound ripple size lies in a window from 5 to 12 Å. When temperature increases, the ripple density shows a similar trend as a function of $D$.

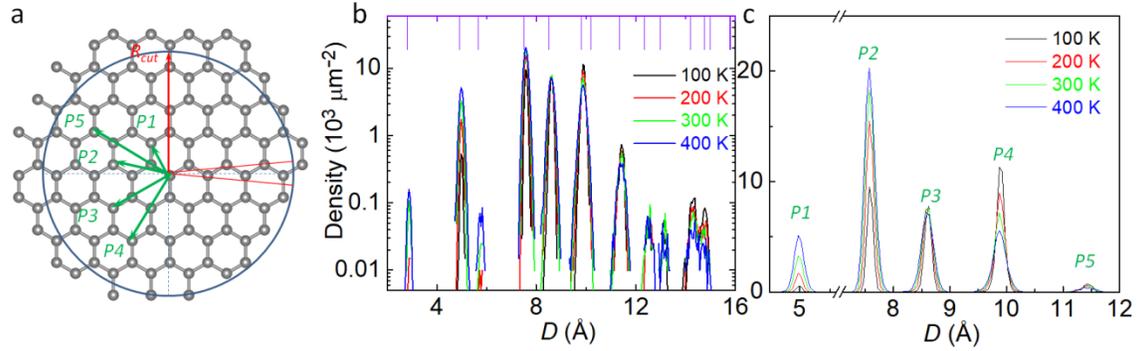

**Figure 1** Ripple distribution in relaxed graphene membrane with the circle-like model. (a) Top view of a circle-like ripple with a carbon atom in the center. (b) Statistical results of the circle-like ripples in relaxed graphene membrane. Purple bar indicates possible values of $D$ ($|m\mathbf{a}+n\mathbf{b}|$) in pristine graphene. (c) Temperature dependence on the distribution of small ripples; P1-P5 indicate the five strongest peaks in (a).

To understand better the effect of temperature on the ripple distribution, we analyze five typical peaks of ripple density with high intention in Figure 1c, which correspond to the five configurations indicated from P1 to P5 in Figure 1a. Our result shows that the width of $D$ peaks becomes broader when the temperature increases, which is reasonable as a higher temperature introduces larger thermal fluctuations. However, not all the ripple density increases as the temperature increases, i.e., for ripples with $D < 8$ Å, the density of ripple increases as $T$ increases; for ripples with $D > 9$ Å, the trend is reversed, while the density of ripple with $D$ around 8.7 Å remains the same.

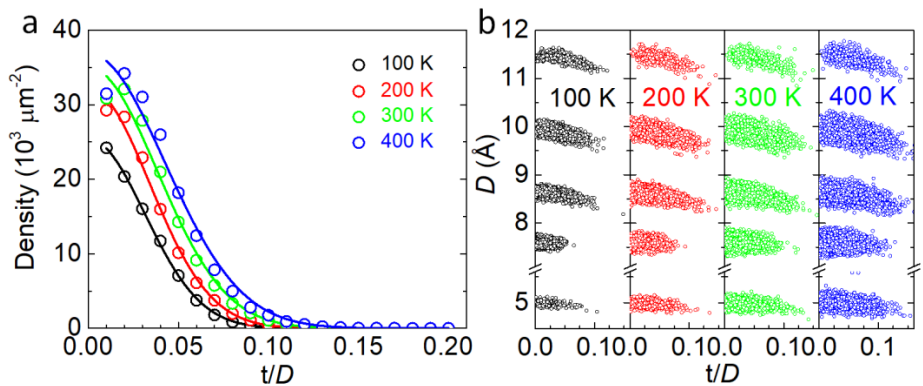

**Figure 2** Ripple density (a) and size *D* (b) as a function of *t/D* in graphene membrane with the circle-like model. The circles are statistical data from MD simulations and the solid lines are fit results using Eq. (2). Black, red, green and blue represent for 100 K, 200 K, 300 K and 400 K, respectively.

Thus, it is hard to describe the ripple distribution only by *D*, therefore we plot the ripple density as a function of *t/D* in Figure 2a. Our result shows that as *T* increases from 100 K to 400 K, the threshold value of *t/D* increases from 0.09 to 0.12. We attribute this to the thermal fluctuation of height *t* under higher temperature. More interestingly, the ripple density decreases as *t/D* increases, indicating that ripples prefer to form a flatter profile instead of a sharper one. Now let us compare these numerical results with our theoretical predictions. Indeed, the ripple sizes counted with the circle-like model are all less than the estimated $L_{th}$. Furthermore, by fitting those data with an exponential function, we found that the ripple distribution is in a good agreement with Eq. (2) for all the temperatures considered in our simulation [see solid line in Figure 3a]. This is not obvious at all since the estimate (2) looks too simple (in particular, it does not take into account interactions between ripples, their slowly decaying correlation functions [1,15,16] and other complications. Nevertheless, this simple estimate turns out to be qualitatively correct. In Table I, we list the fitting parameters in Eq. (2) at different temperatures with an introduced proportionality *A*. For example, at room temperature (*T*=300K), the obtained value of *C* is about 8.63. When the temperature increases, both *A* and *C* increase, which probably reflects the temperature dependence of the bending rigidity *κ* found in the previous study [12], not considered in our fitting process. On the other hand, we notice that when the temperature is over 200 K, the ripple density first increases and then decreases. To clarify this problem, we plot the relation between *D* and *t/D* in Figure 2b. When the ripple size is smaller than 10.5 Å, all ripple density decreases as *t/D* increases; but for ripples around 11.5 Å, the ripple density decreases after the initial increase. This indicates that as the environment temperature increases, more and more large ripples appear in graphene, and finally become dominant above the room temperature. These subtleties go beyond our simple analytical estimate.

**Table I** Fitting parameters of the ripple distribution in Eq. (1) with the circle-like model. Here the fitting is performed within an accuracy of $10^{-4}$ for the values of $P(t, D)$, and $\kappa$ is fixed by using the value of pristine graphene at room temperature, i.e., $\kappa/k_B T = 40$ if T=300K.

| T | 100 K | 200 K | 300 K | 400 K |
|---|---|---|---|---|
| A | 25.39 | 32.20 | 34.99 | 36.86 |
| C | 4.27 | 7.27 | 8.63 | 9.32 |

Now we make an analysis of the simulation data using the arbitrary-shape model. We follow similar steps as performed in the circle-like model, i.e., we calculate {R} in 180 sectors as in the circle-like model, then extract {D} by summing R in each pair of sectors ($\theta$ and $\theta$+180°). In this way, we will have 90 values of $D = R_\theta + R_{\theta+180°}$, forming a set {D} to characterize the size *D* of one ripple. A complete statistics of all 90 values may not be necessary, and here, we evaluate three important values from {D} : $D_{min}$, $D_{max}$ and $D_{ave}$, which stand for the minimum, maximum and average of the 90 values in {D}, respectively. The ripple distributions based on $D_{min}$, $D_{max}$ and $D_{ave}$ are plotted in the left panel of Figure 3. Clearly, three *D* values obtained using the arbitrary-shape model are all much larger than those obtained using the circle-like model, especially $D_{max}$ and $D_{ave}$. When the ripple size is characterized by $D_{min}$, the ripple density is discrete within the range from 4 to 17 Å and the highest peak appears around 4 Å, while the same data analyzed by $D_{max}$ shows that the ripples accumulate from 1.5 to 3.7 nm with the highest peak around 3.45 nm. On the contrary, the peaks obtained by $D_{ave}$ are nearly continuously distributed, and those ripples are highly accumulated in the range from 1.8 to 2.5 nm.

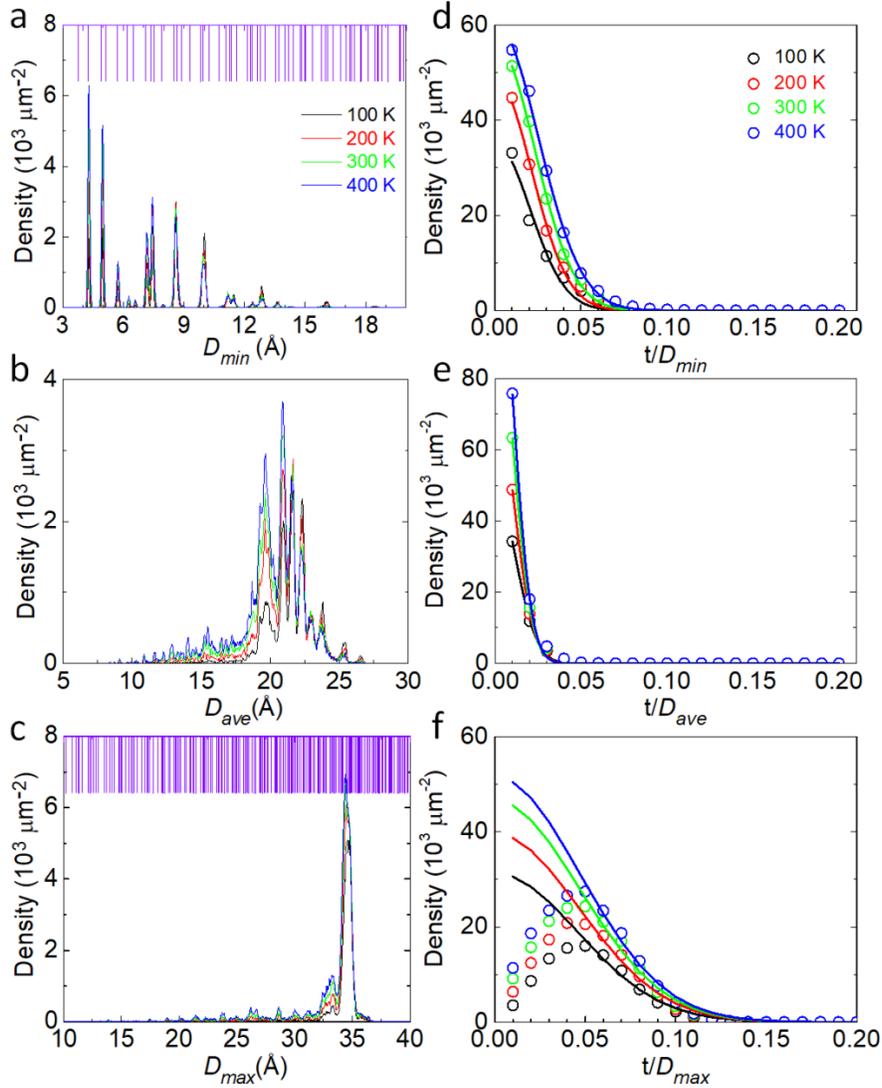

**Figure 3** Temperature-dependent ripple distribution in graphene membrane using the arbitrary-shape model. The left panel shows the ripple density as a function of $D_{min}$, $D_{max}$ and $D_{ave}$, respectively. The purple bar indicates possible ripple size using the lattice of pristine graphene. The right panels show the corresponding relations between the ripple density and t/D using $D_{min}$, $D_{max}$ and $D_{ave}$.

Now, we turn to the relation between the ripple density and *t/D* as shown in the right panels of Figure 3. In general, the numerical results using $D_{min}$, $D_{max}$ and $D_{ave}$ show similar dependence of $P(t,D)$ just as in the circle-like model, but with a much faster decreasing rate. When *t/D$_{min}$* (*t/D$_{ave}$*) is larger than

0.08 (0.06), the density of ripple becomes zero, this is because the statistic $D$ values here are much larger than those in the circle-like model, resulting in a smaller threshold of $t/D$. However, if the ripple density is counted by $t/D_{max}$, we see that it first increases to a maximum value around $t/D_{max} = 0.05$, and then decreases gradually to 0 around $t/D_{max} = 0.12$, which is close to the threshold of the circle-like model. In fact, as a ripple is generally anisotropic and $D_{max}$ is the maximum value from $R_\theta + R_{\theta+180°}$, there could be cases that the decay of the ripple is very slow along one direction, therefore $t$ will keep small, and $D_{max}$ could be very large, leading to small $t/D_{max}$. Therefore, $D_{max}$ may not be a good quantity to characterize the real ripple size, especially if we obtain a small $t/D_{max}$. Once again, by performing fitting processes with the data based on $D_{min}$ or $D_{ave}$, we find that the ripple distribution agrees well with the prediction in Eq. (2). The corresponding fitting parameters of $A$ and $C$ are listed in Table II. However, it is notable that the value of $C$ using $D_{min}$ or $D_{ave}$ is one or two orders larger than that in the circle-like model. In the case of $D_{max}$, because the value of $D_{max}$ is accumulated around 3.45 nm which is larger than the estimated $L_{th} \approx 3\ nm$, we have to fit the data by Eq. (5) instead of Eq. (2). Due to the reason mentioned above, $P(t, D)$ can only be captured by the theoretical prediction for large $t/D_{max}$, namely, the ripple distribution using the definition of $D_{max}$ follows the expression of Eq. (5). Besides, the values of $C$ at different temperatures are all in the same order as that in the ripple-like model. To this end, we can conclude that, for a given graphene membrane, if the definition of the ripple size is different, then the distribution function of ripples may vary dramatically. Even so, all of them meet the standard or corrected Boltzmann distribution.

**Table II** Fitting parameters of the ripple distribution using the arbitrary-shape model. Here Eq. (2) is adopted for data based on $D_{min}$ and $D_{ave}$, and Eq. (5) is used for $D_{max}$.

|  | $t/D_{min}$ | | $t/D_{ave}$ | | $t/D_{max}$ | |
| --- | --- | --- | --- | --- | --- | --- |
| T | A | C | A | C | A | C |
| 100 K | 35.2 | 9.94 | 47.76 | 28.12 | 31.30 | 1.78 |
| 200 K | 48.93 | 18.51 | 72.89 | 67.40 | 39.60 | 3.45 |
| 300 K | 56.52 | 23.68 | 99.14 | 112.79 | 46.61 | 5.17 |
| 400 K | 60.54 | 26.39 | 120.57 | 156.01 | 51.62 | 6.78 |

In summary, we provide a theoretical analysis of the ripple distribution in graphene membrane at finite temperature and verified our estimations via the semiclassical molecular dynamics simulations. We introduced two geometric models to study numerically the distribution of ripples in simulated graphene membrane. In the circle-like model, we emphasize the effect of temperature on the density of ripples. In a contrast to some intuitive feeling, not all the density of ripples will increase when the ambient temperature increases. In the arbitrary-shape model, we conclude that if the ripple size is defined differently, one can get totally different distribution functions for the same membrane. More importantly, we found in both models that once the ripple size is well defined, the ripple density decays exponentially as the curvature ($t/D$) increases in a way of the standard or corrected Boltzmann distribution for small and large ripples, respectively. These numerical results are constant with our theoretical predications derived analytically from the bending energy of the membrane. Thus, our combined theoretical and numerical studies will present a deep understanding of ripples in graphene and other related 2D materials.

## Acknowledgments

This work was supported by the National Key R&D Program of China (Grant No. 2018FYA0305800), the program for professor of special appointment (Eastern Scholar) at Shanghai Institutions of higher learning and by Dutch Science Foundation NWO/FOM under Grant No. 16PR1024. M.I.K. thanks financial support from JTC-FLAGERA Project GRANSPORT. Support by the Netherlands National Computing Facilities Foundation (NCF) with funding from the Netherlands Organization for Scientific Research (NWO) is gratefully acknowledged.

**Corresponding author**: s.yuan@whu.edu.cn

## References

(1) D. R. Nelson, T. Piran and S. Weinberg, Statistical Mechanics of Membranes and Surfaces. (World Scientific, Singapore) (**2004**).

(2) D. R. Nelson, L. Peliti, Fluctuations in membranes with crystalline and hexatic order. *J. Physique* **1987,** *48(7),* 1085.

(3) J. A. Aronovitz, T. C. Lubensky, Fluctuations of solid membranes. *Phys. Rev. Lett.* **1988,** *60,* 2634.

(4) M. J. Bowick, A. Travesset, The statistical mechanics of membranes. *Phys. Rep*. **2001,** *344,* 255.

(5) P. Le Doussal, L. Radzihovsky, Self-consistent theory of polymerized membranes. *Phys. Rev. Lett.* **1992,** *69,* 1209.

(6) D. Gazit, Structure of physical crystalline membranes within the self-consistent screening approximation. *Phys. Rev. E* **2009,** *80,* 041117.

(7) J. P. Kownacki, D. Mouhanna, Crumpling transition and flat phase of polymerized phantom membranes. *Phys. Rev. E* **2009,** *79,* 040101(R).

(8) P. Le Doussal, L. Radzihovsky, Anomalous elasticity, fluctuations and disorder in elastic membranes. *Ann. Phys.* **2018,** *392,* 340.

(9) A. Mauri and M. I. Katsnelson. Scaling behavior of crystalline membranes: An ε-expansion approach. *Nucl. Phys. B* **2020,** *956,* 115040.

(10) K. S. Novoselov, A. K. Geim, S. V. Morozov, D. Jiang, Y. Zhang, S. V. Dubonos, I. V. Grigorieva, A. A. Firsov, Electric Field Effect in Atomically Thin Carbon Films. *Science* **2004,** *306(5696),* 666.

(11) J. C. Meyer, A. K. Geim, M. I. Katsnelson, K. S. Novoselov, T. J. Booth & S. Roth, The structure of suspended graphene sheets. *Nature* **2007,** *446,* 60.

(12) A. Fasolino, J. H. Los & M. I. Katsnelson, Intrinsic ripples in graphene. *Nature Mater.* **2007,** *6,* 858.

(13) C. H. Lui, L. Liu, K. F. Mak, G. W. Flynn, T. F. Heinz, Ultraflat graphene. *Nature* **2009,** *462(7271),* 339-341.

(14) A. L. Vázquez de Parga, F. Calleja, B. Borca, M. C. G. Passeggi, J. J. Hinarejos, F. Guinea, R. Miranda, Periodically Rippled Graphene: Growth and Spatially Resolved Electronic Structure. *Phys. Rev. Lett.* **2008,** *100(5),* 056807.

(15) M. I. Katsnelson, The physics of graphene, 2nd Edition, Cambridge University Press, **2020**, Chapter 9.

(16) M. I. Katsnelson & A. Fasolino, Graphene as a prototype crystalline membrane. *Accounts Chem. Research* **2013,** *46,* 97.

(17) V. B. Shenoy, C. D. Reddy, A. Ramasubramaniam, Y. W. Zhang, Edge-Stress-Induced Warping of Graphene Sheets and Nanoribbons. *Phys. Rev. Lett.* **2008,** *101(24)*, 245501.

(18) M. I. Katsnelson & A. K. Geim, Electron scattering on microscopic corrugations in graphene. *Phil. Trans. R. Soc. A* **2008,** *366,* 195.

(19) D. W. Boukhvalov, & M. I. Katsnelson, Enhancement of chemical acitivty in corrugated graphene. *J. Phys. Chem. C* **2009,** *113,* 14176.

(20) A. Košmrlj, D. R. Nelson. Response of thermalized ribbons to pulling and bending. *Phys. Rev. B* **2016,** *93,* 125431.

(21) J. H. Los, A. Fasolino and M. I. Katsnelson, Scaling behavior and strain dependence of in-plane elastic properties of graphene. *Phys. Rev. Lett.* **2016**, *116***,** 015901.

(22) J. H. Los, A. Fasolino and M. I. Katsnelson. Mechanics of thermally fluctuating membranes. *NPJ 2D Mater. Appl.* **2017,** *1***,** 9.

(23) P. Z. Sun, Q. Yang, W. J. Kuang, Y. V. Stebunov, W. Q. Xiong, J. Yu, R. R. Nair, M. I. Katsnelson, S. J. Yuan, I. V. Grigorieva, M. Lozada-Hidalgo, F. C. Wang, A. K. Geim, Limits on gas impermeability of graphene. *Nature* **2020,** *579(7798)*, 229-232.

(24) L. D. Landau & E. M. Lifshitz, *Theory of Elasticity*. Oxford: Pergamon. **1970**.

(25) S. J. Plimpton, *J. Comput. Phys.* **1995,** *117*, 1.